\documentclass[prd,aps,preprint,showpacs,nofootinbib]{revtex4}
\usepackage{epsfig}
\usepackage{amsmath}

\begin{document}

%\preprint{LBNL-xxx}\preprint{RBRC-xxx}

\title{Single Spin Asymmetry in Inclusive Hadron Production in $pp$
Scattering from Collins Mechanism}

\author{Feng Yuan}
\email{fyuan@lbl.gov} \affiliation{Nuclear Science Division,
Lawrence Berkeley National Laboratory, Berkeley, CA
94720}\affiliation{RIKEN BNL Research Center, Building 510A,
Brookhaven National Laboratory, Upton, NY 11973}
%\date{\today}

\begin{abstract}
We study the Collins mechanism contribution to the single transverse
spin asymmetry in inclusive hadron production in $pp$ scattering
$p^{\uparrow}p\to \pi X$ from the leading jet fragmentation.
The azimuthal asymmetric distribution
of hadron in the jet leads to a single spin asymmetry for the produced
hadron in the Lab frame. The effect is evaluated in a transverse momentum
dependent model that takes into account the transverse momentum
dependence in the fragmentation process.
We find the asymmetry is comparable in size to the experimental observation
at RHIC at $\sqrt{s}=200 GeV$.
\end{abstract}

\maketitle

\newcommand{\be}{\begin{equation}}
\newcommand{\ee}{\end{equation}}
\newcommand{\ben}{\[}
\newcommand{\een}{\]}
\newcommand{\beqn}{\begin{eqnarray}}
\newcommand{\eeqn}{\end{eqnarray}}
\newcommand{\Tr}{{\rm Tr} }

Single-transverse spin asymmetries (SSA) in hadronic processes have a long history \cite{E704,Bunce}.
They are defined as the spin asymmetries when we flip the transverse spin
of one of the hadrons involved in the scattering:
$A=(d\sigma(S_\perp)-d\sigma(-S_\perp))/(d\sigma(S_\perp)+d\sigma(-S_\perp))$.
Recent experimental studies of SSAs in polarized semi-inclusive lepton-nucleon
deep inelastic scattering (SIDIS) \cite{hermes,dis}, in hadronic
collisions \cite{star,star2,phenix,brahms,bra2}, and in the relevant
$e^+e^-$ annihilation process \cite{belle}, have renewed the
theoretical interest in SSAs and in understanding their roles in
hadron structure and Quantum Chromodynamics (QCD).
Among the theoretical approaches proposed in the QCD framework,
the transverse momentum dependent (TMD) parton distribution approach
\cite{Siv90,Col93,Ans94,MulTan96,BroHwaSch02,Col02,BelJiYua02,BoeMulPij03}
and the twist-3 quark-gluon correlation approach \cite{Efremov,qiu} are the most
discussed in the last few years, and it has been demonstrated that these
two approaches are actually consistent with each other in the overlap regions where
both apply~\cite{Ji:2006ub}.

For the SSAs in hadron production, two important contributions have been
identified in the literature: one is associated
with the so-called Sivers effect~\cite{Siv90,qiu} from the incoming polarized
nucleon; and one with the Collins effect~\cite{Col93} in the fragmentation
process for the final state hadron. A third contribution associated with the
twist-three effects from the unpolarized nucleon was found very 
small~\cite{Kanazawa:2000hz}. Both effects shall contribute to the
SSA in inclusive hadron production in nucleon-nucleon scattering,
for example, in pion production in $p^{\uparrow}p\to\pi X$.
However, how the transverse momentum dependent Sivers and Collins functions contribute to the inclusive hadron production in $p^{\uparrow}p\to\pi X$ is not clear, because the large $p_\perp$
of the final state hadron has no direct connection with the intrinsic transverse momentum
of partons in nucleon or the transverse momentum in the fragmentation
process. Therefore, these effects can only be evaluated
in a model-dependent way~\cite{Ans94,{Anselmino:2005sh}}. Meanwhile, for the Sivers effects, the initial
and final state interactions are crucial to the nonzero SSA in
the hadronic processes~\cite{BroHwaSch02}, which have not
yet been implemented in the model calculations~\cite{Ans94,{Anselmino:2005sh}}.
Thus, it is more appropriate to adopt the twist-3
quark-gluon correlation approach for the Sivers contribution in
$p^\uparrow p\to \pi X$, which takes into account
the initial and final state interaction effects in the formalism~\cite{qiu}.

For the Collins effect, a twist-3 extension to the
fragmentation process has been formulated in~\cite{Koike:2002ti}.
However, a universality argument for the Collins function~\cite{Metz:2002iz}
would indicate the
contributions calculated in~\cite{Koike:2002ti} vanishes. This universality
has also been recently extended to $pp$ collisions
\cite{collins-s}. Therefore, to establish a consistent framework for the  twist-three
quark-gluon correlation contribution in the fragmentation process,
we need further theoretical developments. Before that, it is worthwhile to
investigate the Collins effects contribution to the inclusive hadron's SSA
$p^{\uparrow}p\to\pi X$ by extending the results of \cite{collins-s} and using a transverse
momentum dependent model in the quark fragmentation. This is what we will explore in this paper. Earlier
works on the Collins contribution can be found in~\cite{Anselmino:1999pw,Boglione:1999dq,{Anselmino:2004ky}}.

In our model, we assume a transversely polarized quark is produced in hard
partonic processes with transverse momentum $P_\perp$ and rapidity $y_1$.
This transversely polarized quark then fragments into a final state hadron
with azimuthal asymmetric distribution (relative to the jet) according to
the Collins effect~\cite{collins-s}. The final state hadron's momentum
will naturally be the jet's momentum in a fraction of $z_h$ plus the fragmentation
momentum of hadron relative to the jet: $P_{hT}$. Thus, the
azimuthal asymmetry found in \cite{collins-s} will lead to
an azimuthal asymmetry of final state hadron in the Lab frame, which
is exactly the experimental measurement of the left-right asymmetry $A_N$.
Our approach is a semi-classic picture, in the sense that the quark jet production
comes from the hard partonic processes and is calculated from a collinear
factorization approach, whereas the fragmentation process takes the TMD
effects explicitly. This assumption, of course, will introduce some theoretical
uncertainties. However, we argue that our results shall provide a good estimate
on how large the Collins  effects contribute to the inclusive hadron's
SSA in $p^{\uparrow}p\to\pi X$.
It is important to note that, to make reliable predictions for the inclusive process in $pp $ scattering
at the transverse momentum region
of our interest, we have to take into account the high order perturbative resummation
corrections, and the power corrections as well~\cite{Laenen:2000de}. It will be interested
to study how this affect the spin asymmetries we are exploring in this paper, as well
as other spin-dependent observables.

There have been calculations for the Collins effects contributions to inclusive hadron's SSA
$p^{\uparrow}p\to\pi X$ in the transverse momentum dependent approach
similar to our model, where the Collins contribution was found strongly
suppressed~\cite{Anselmino:2004ky}\footnote{After some corrections, this approach also predicts sizable
contribution from the Collins effects~\cite{newmodel}.}.
However, from the following calculations,
we will find that the contributions are as large as the SSAs observed by the RHIC experiments
at $\sqrt{s}=200GeV$.
\begin{figure}[t]
\begin{center}
\includegraphics[width=9cm]{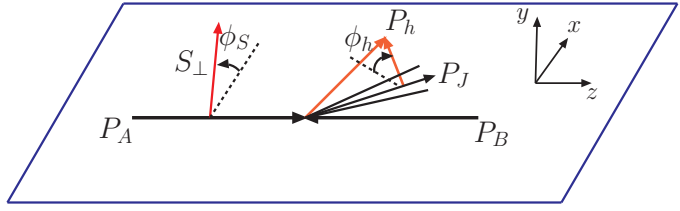}
\end{center}
\vskip -0.4cm \caption{\it Illustration of the kinematics for the
azimuthal distribution of hadrons in a leading jet fragmentation in single
transversely polarized $pp$ scattering.} \label{fig1}
\end{figure}
In our model, we will follow the Collins effects in a jet fragmentation in single transversely
polarized $pp$ scattering~\cite{collins-s}.
As illustrated in Fig.~1, we studied the process,
\begin{equation}
p(P_A,S_\perp)+p(P_B)\to jet (P_J)+X\to H(P_h)+X\ , \label{e1}
\end{equation}
where a transversely polarized proton with momentum $P_A$ scatters
on another proton with momentum $P_B$, and produces a jet with
momentum $P_J$ (transverse momentum $P_\perp$ and rapidity $y_1$
in the Lab frame). The three momenta of $P_A$, $P_B$ and $P_J$
form the so-called reaction plane. Inside the produced jet,
the hadrons are distributed around the jet axes.
A particular hadron $H$ will carry certain longitudinal momentum
fraction $z_h$ of the jet, and its transverse momentum $P_{hT}$ relative
to the jet axis will define an azimuthal angle with the reaction
plane: $\phi_h$, shown in Fig.~1. Thus, the hadron's momentum
is defined as $\vec{P}_h=z_h \vec{P}_J+\vec{P}_{hT}$. The relative transverse momentum
$P_{hT}$ is orthogonal to the jet's momentum $P_J$: $\vec{P}_{hT}\cdot \vec{P}_J=0$.
Similarly, we can define the
azimuthal angle of the transverse polarization vector of the
incident polarized proton: $\phi_S$.
The Collins effect will contribute to an azimuthal asymmetry
for hadron production in term of $\sin(\phi_h-\phi_S)$. The
differential cross section can be written as~\cite{collins-s}
\begin{eqnarray}
\frac{d\sigma}{dy_1dy_2dP_\perp^2dzd^2P_{hT}}=\frac{d\sigma}{d{\cal
P.S.}}=\frac{d\sigma_{UU}}{d{\cal P.S.}}+
|S_\perp|\frac{|P_{hT}|}{M_h}\sin(\phi_h-\phi_s)
\frac{d\sigma_{TU}}{d{\cal P.S.}}\ ,\label{e8}
\end{eqnarray}
where $d{\cal P.S.}={dy_1dy_2dP_\perp^2dzd^2P_{hT}}$ represents
the phase space for this process, $y_1$ and $y_2$ are rapidities
for the jet $P_J$ and the balancing jet, respectively, $P_\perp$
is the jet transverse momentum, and the final observed hadron's
kinematic variables $z_h$ and $P_{hT}$ are defined above.
$d\sigma_{UU}$ and $d\sigma_{TU}$ are the the spin-averaged
and single-transverse-spin dependent cross section terms,
respectively. They are defined as~\cite{collins-s}
\begin{eqnarray}
\frac{d\sigma_{UU}}{d{\cal P.S.}}&=&\sum_{a,b,c}x'f_{b}(x')xf_a(x)
D_c^h(z,P_{hT})H_{ab\to
cd}^{uu} \ ,\nonumber\\
\frac{d\sigma_{TU}}{d{\cal P.S.}}&=&\sum_{b,q}x'f_{b}(x')x\delta
q_T(x) \delta \hat q(z,P_{hT})H_{qb\to qb}^{\rm Collins} \ .
\end{eqnarray}
Here, $x$ and $x'$ are the momentum fractions carried by the parton
``$a$" and``$b$" from the incident hadrons, respectively. In the above
equation, $f_a$ and $f_b$ are the associated parton distributions,
$D_q(z_h,P_{hT})$ is the TMD quark fragmentation function,
$\delta q_T(x)$  is the quark transversity distribution, and $\delta
\hat q(z_h,P_{hT})$ the Collins fragmentation function.
The hard factors for the spin-averaged cross sections are
identical to the differential partonic cross sections: $H_{ab\to
cd}^{uu}=d\hat\sigma_{ab\to cd}^{uu}/d\hat t$, and the spin-dependent
hard factors have been calculated in~\cite{collins-s}.

In the following, we will study how the above azimuthal asymmetry contributes
to the SSA in inclusive hadron production in $pp$
scattering $p^{\uparrow}p\to \pi X$, especially at RHIC energy. In order to estimate this contribution,
we assume that the hadron production is dominated by the leading jet
fragmentation, and the Collins effects discussed above shall lead to a
nonzero azimuthal asymmetry in the Lab frame, for example, in term of
$\sin(\Phi_h-\Phi_S)$ where $\Phi_h$ and $\Phi_S$ are the azimuthal angles of
the final state hadron and the polarization vector in the Lab frame. Following this assumption, the hadron production
follows two steps: jet production and hadron fragmentation. In the fragmentation process,
as we mentioned above, the hadron's momentum $\vec{P}_h$ will be
\begin{equation}
\vec{P}_h=z_h \vec{P}_J+\vec{P}_{hT}\ .
\end{equation}
If we choose
the jet transverse momentum direction as $\hat x$ direction as we plotted in Fig.~1,
the final hadron's momentum can be parameterized as follows,
\begin{eqnarray}
P_{hx}&=&z_h P_\perp+P_{hT}\cos \phi_h\cos\theta\ , \nonumber\\
P_{hy}&=&P_{hT}\sin\phi_h\ ,\nonumber\\
P_{hz}&=&z_h P_{Jz}-P_{hT}\cos\phi_h\sin\theta \ ,
\end{eqnarray}
where $P_\perp$ is the transverse momentum of the jet in the Lab frame,
$\theta$ the polar angle between the jet and plus $\hat z$ direction (the polarized
nucleon momentum direction): $y_1\approx \eta=-\ln\tan(\theta/2)$, and
$y_1$ and $\eta$ are the rapidity and pseudorapidity of the hadron, respectively.
We can also work out the general results for any azimuthal angle ($\Phi_J$) of
the jet in the Lab frame.
At RHIC experiment, a sizable single spin asymmetry
has been observed in the forward direction, which means
$\theta\approx 0$. We further assume that $P_{hT}\ll P_\perp$, so that
the rapidity of the hadron will approximately equal to the jet's rapidity. The uncertainties
coming from this approximation can be further studied by taking into
account the full kinematics in the fragmentation process.
With the above kinematics of $P_{hx}$, $P_{hy}$, and $P_{hz}$, we will
be able to derive the transverse momentum $P_{h\perp}$ and azimuthal
angle $\Phi_h$ for the final state hadron in the Lab frame.

By integrating the fragmentation functions over $z_h$ and $P_{hT}$, we
will obtain the differential cross sections and the spin asymmetries depending on
the final state hadron's kinematics, $y_1$ and $\vec{P}_{h\perp}$, where
$\vec{P}_{h\perp}$ is hadron's transverse momentum in the Lab
frame. Let us first estimate roughly how the above effect contributes to the SSA for
pion production in $pp$ scattering $p^\uparrow p\to \pi X$, especially for the sign.
Suppose the incident nucleon $A$ is polarized
along the $\hat y$ direction, and we assume that $\pi^+$ is dominated by the
valence $u$-quark fragmentation in the forward rapidity region. The HERMES data
show that the Collins function for $u$-quark fragmentation into $\pi^+$
is negative if the $u$-quark transversity distribution is positive in the valence region~\cite{Vogelsang:2005cs}.
From the differential cross section Eq.~(\ref{e8}), we will find that the $\pi^+$ will
prefer to be produced with $\phi_h$ around $0$, which will lead to an increase of
$\pi^+$ production in $+\hat x$ direction. That means this contribution will result into
a positive left-right ($A_N$) asymmetry for $\pi^+$. Similarly, we find that the contribution
to $\pi^-$ left-right asymmetry is negative, and that for $\pi^0$ will be determined by
the contributions from both $u$ and $d$ quarks. These estimates are consistent with
the experimental trends for the SSAs in pion productions at RHIC~\cite{star,star2,{brahms},bra2}.

Quantitatively, we can perform our calculations for the spin asymmetries at RHIC
energy. With the above kinematics, we can write down the differential cross section
for inclusive hadron production $pp\to \pi X$ coming from the leading jet fragmentation,
depending on the final state hadron's kinematics,
\begin{eqnarray}
\frac{d\sigma^{uu}}{dy_1d^2P_{h\perp}}&=&\int dy_2 dP_\perp^2\frac{1}{\pi}d\Phi_J dz_h
\Theta(P_\perp-{\bf{k}}_0)\Theta(\Lambda-P_{hT})\nonumber\\
&&\times
xf_a(x)x'f_b(x')D_c(z_h,P_{hT})H^{uu} \ ,
\end{eqnarray}
where the jet's transverse momentum $\vec{P}_\perp$ is integrated out, and also the
associated azimuthal angle $\Phi_J$. From the rotation invariance of the above expression,
the differential cross section will be azimuthal symmetric for the final state hadron. Thus,
it will not depend on the azimuthal angle $\Phi_h$.  In the above equation, we
have imposed two cuts for the momenta $P_\perp$ and $P_{hT}$. The minimum
value for $P_\perp$ is necessary to guarantee that the fragmentation is coming from the
leading jet production, whereas a cut on $P_{hT}$ is needed to ensure that we will not get into
un-physical region in the fragmentation process. Theoretical uncertainties can be further
studied by varying these two parameters.
Similarly the spin dependent cross section can be written as
\begin{eqnarray}
\frac{d\Delta\sigma^{UT}(S_\perp)}{dy_1d^2P_{h\perp}}&=&\int dy_2k_\perp dP_\perp\frac{1}{\pi}d\Phi_J dz_h
\Theta(P_\perp-{\bf{k}}_0)\Theta(\Lambda-P_{hT})\nonumber\\
&&\times\frac{|P_{hT}|}{M_h}\sin(\phi_h-\Phi_S+\Phi_J)
x\delta q_T(x)x'f_b(x') \delta \hat q(z_h,P_{hT})H^{\rm Collins}_{qb\to qb} \ ,
\end{eqnarray}
where $\Phi_S$ is the azimuthal angle of the transverse polarization
vector $S_\perp$ in the Lab frame, and its relative angle to the jet defined in
Fig.~1 $\phi_S$ can be written as $\phi_s=\Phi_S-\Phi_J$.
Following above, we further define $\Phi_h$ as
the azimuthal angle of the produced hadron in the Lab frame, which
is different from the above $\phi_h$.
From these differential cross sections, the left-right asymmetry $A_N$ is calculated as
\begin{equation}
A_N=\frac{\langle 2\sin(\Phi_S-\Phi_h)d\Delta\sigma^{UT}\rangle}{\langle d\sigma^{uu}\rangle} \ .
\label{colasy}
\end{equation}
In the numerical simulations, we use simple Gaussian parameterizations for the TMD fragmentation
functions,
\begin{eqnarray}
D_c(z_h,P_{hT})&=&\frac{1}{\pi \langle p_\perp^2\rangle}e^{-P_{hT}^2/\langle p_\perp^2\rangle}D_c(z_h) \ , \nonumber\\
\delta\hat q(z_h,P_{hT})&=&\frac{2M_h}{(\pi\langle p_\perp^2\rangle)^{3/2}}
e^{-P_{hT}^2/\langle p_\perp^2\rangle}\delta \hat q^{(1/2)}(z_h) \ ,
\end{eqnarray}
where $D_c(z_h)$ is the integrated fragmentation function, and $\delta \hat q^{(1/2)}$ the so-called
half-moment of the Collins function. The above parameterizations have been chosen
to give the right normalization for the two fragmentation functions. In the
following numerical calculations, we choose the parameters $\langle p_\perp^2\rangle=0.2 GeV^2$,
$\Lambda=1GeV$, and ${\bf k}_0=1GeV$.

The half-moment of the Collins functions $\delta\hat q^{(1/2)}(z_h)$ have been determined
from the HERMES data by assuming some functional form dependence on $z_h$~\cite{Vogelsang:2005cs,Efremov:2006qm,{Anselmino:2007fs}}.
In \cite{Vogelsang:2005cs}, they are parameterized as $\delta \hat q^{(1/2)}=Cz_h(1-z_h)D_c(z_h)$
for the favored and unfavored ones. These parameterizations have to be updated, because
the di-hadron production in $e^+e^-$ annihilation from BELLE experiments showed
a strong increase of the asymmetry with $z_h$~\cite{belle}. To be consistent with this observation,
we re-parameterize the Collins functions as follows~\cite{Efremov:2006qm},
\begin{eqnarray}
\delta \hat q_{fav.}^{\pi(1/2)}(z_h)&=&C_f' z_h D_u^{\pi^+}(z_h)\ ,\nonumber\\
\delta \hat q_{unfav.}^{\pi(1/2)}(z_h)&=&C_u' z_h D_d^{\pi^+}(z_h)\ .
\end{eqnarray}
With the new parameters modified from~\cite{Vogelsang:2005cs}: $C_f'=0.61 C_f$ and $C_u'=0.65 C_u$,
we will be able to reproduce the Collins asymmetries for $\pi^\pm$ from HERMES,
assuming the quark transversity distributions follow the parameterizations in~\cite{Martin:1997rz}.

\begin{figure}[t]
\begin{center}
\includegraphics[height=5.0cm,angle=0]{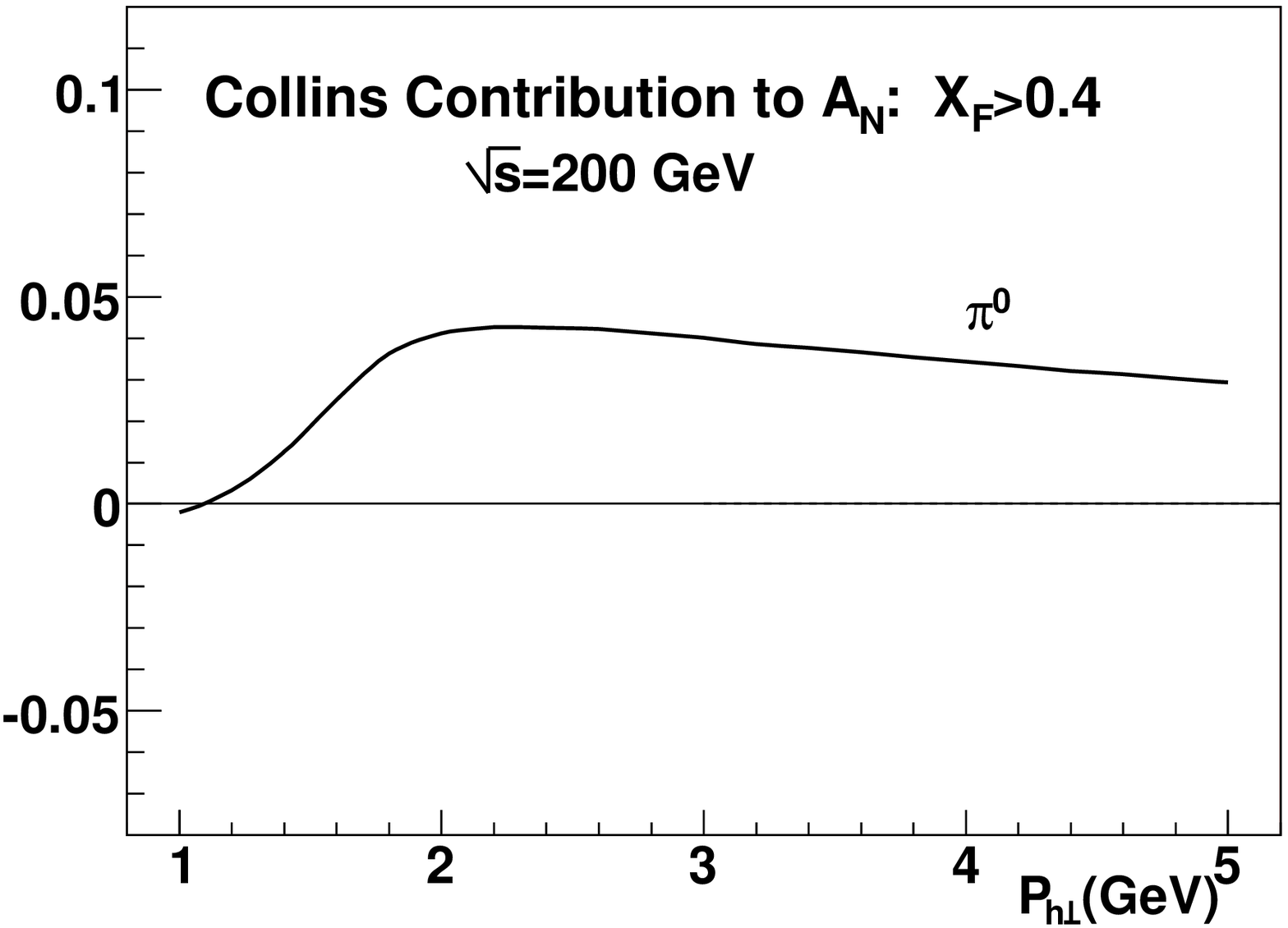}
\includegraphics[height=5.0cm,angle=0]{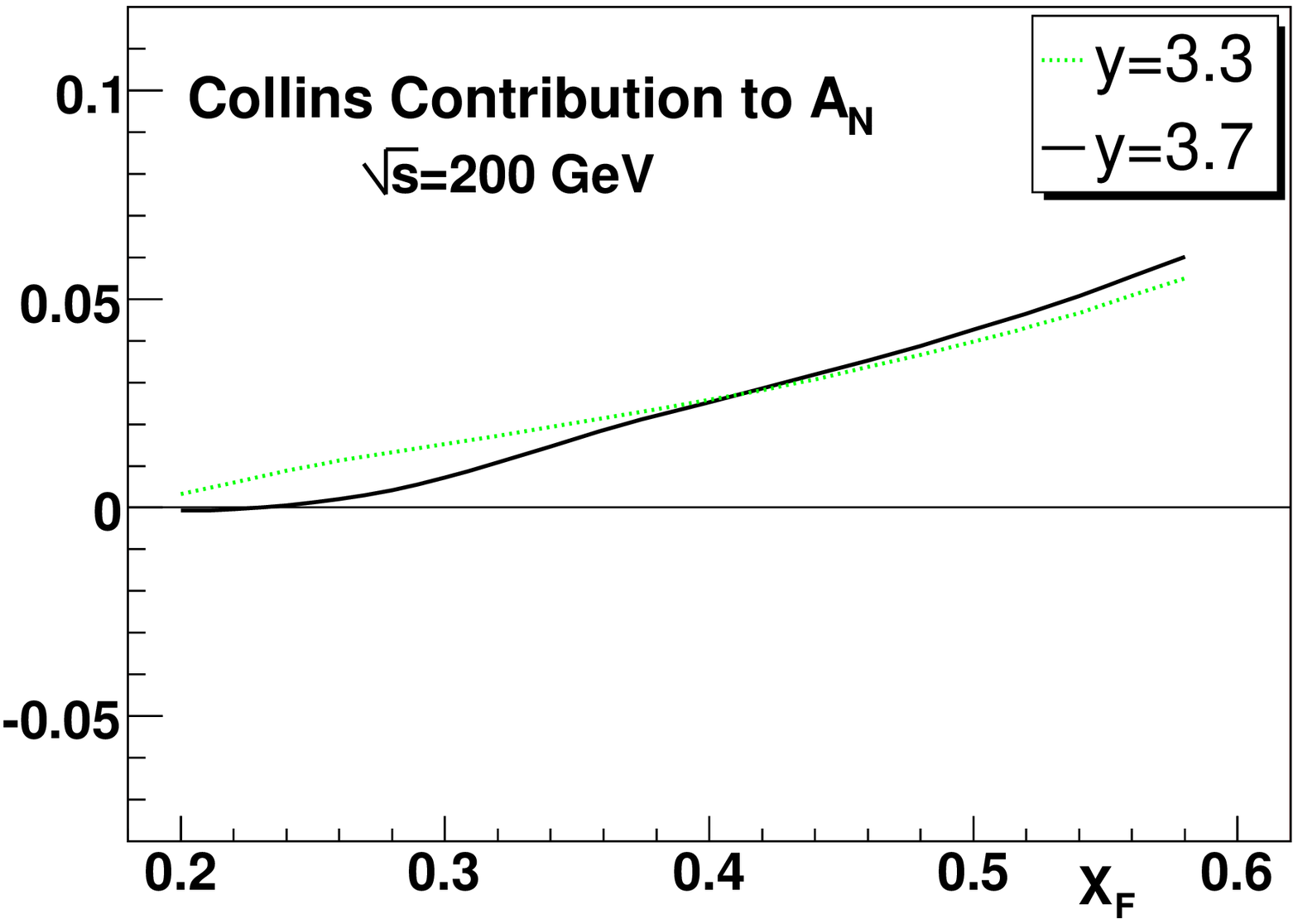}
\vspace*{-0.2cm} \epsfysize=4.2in
\end{center}
\caption{\it Model predictions for the Collins contribution to the SSA in $\pi^0$ production at RHIC
at $\sqrt{s}=200GeV$: left panel as function of $P_\perp$ for
$x_F>0.4$ and integrate over all rapidity; right panel as function of $x_F$ for two different
rapidity bins, $y=3.3,~3.7$, respectively. }
\end{figure}

In Fig.~2, we show the numerical estimates from our model
calculations of the Collins mechanism
contributions to the SSA in $\pi^0$ production in $pp$ scattering  $p^{\uparrow}p\to \pi^0 X$
at RHIC at $\sqrt{s}=200GeV$:
left panel as function of $P_{h\perp}$ for
$x_F>0.4$ and integrate over all rapidity; the right panel as functions of $x_F$ for two different
rapidities: $y=3.3,~3.7$, respectively. Similar results are also obtained for the
charged pions. The decrease of the SSA as $P_{h\perp}$ decreases
is due to our modeling the Collins effects in the fragmentation process. Numerically,
this decrease comes from a lower cut for the jet transverse momentum $P_\perp >{\bf k}_0$
in our formalism Eqs.~(6,7), which limits the Collins contribution to the asymmetries
at $P_{h\perp}$ below the cut-off ${\bf k}_0$. At this kinematic region, however,
a soft mechanism may be responsible for the cross section and
the asymmetries, and its contribution may further change the $P_{h\perp}$ dependence.

These plots show that the Collins contributions to the
SSA in inclusive hadron production in $pp$ scattering $p^\uparrow p\to\pi X$ are not
negligible, rather comparable in size to what we observed at RHIC for charged and
neutral pions~\cite{star,star2,brahms,bra2}. However, we will not intend to compare
them with the real data on these
SSAs, for which we have to take into account the Sivers contributions as well~\cite{qiu}.

In conclusion, in this paper, we have studied the Collins mechanism contribution to the
inclusive hadron's SSA in $pp$ scattering  $p^{\uparrow}p\to \pi X$ at RHIC in a model where the hadron
production comes from the leading jet fragmentation with transverse momentum dependence.
We found that their
contributions to the SSA for inclusive $\pi^0$ production at RHIC are the same size
as the experimental measurements.
Our results shall stimulate more theoretical investigations toward
a fully understanding for the longstanding SSA phenomena
in hadronic processes.

We thank Gerry Bunce, Les Bland, Matthias Grosse-Perdekamp,
Jianwei Qiu, Werner Vogelsang for discussions
and comments. We also thank Mauro Anselmino, Umberto D'Alesio, and Elliot Leader for
the communications concerning the calculations of \cite{Anselmino:2004ky}.
This work was supported in part by the U.S. Department of Energy
under contract DE-AC02-05CH11231. We are grateful to RIKEN,
Brookhaven National Laboratory and the U.S. Department of Energy
(contract number DE-AC02-98CH10886) for providing the facilities
essential for the completion of this work.

\end{document}